\documentclass[11pt]{article}
\usepackage{longtable}
\usepackage{ae,aecompl,pslatex}
\usepackage[dvips]{graphicx}
\usepackage{enumerate,color}
\usepackage{fullpage}
\usepackage{lineno}
\usepackage{setspace}

\usepackage{placeins}
\usepackage{psfrag,epsfig,graphicx,color,array,dcolumn,tabularx,delarray,longtable,indentfirst,amsmath,amsfonts,amssymb,amsthm,eucal,bbm}
\theoremstyle{bkaexa} 
\usepackage{caption}
\usepackage{setspace}
\theoremstyle{bkaexa} 

\theoremstyle{bkathm} 

\theoremstyle{bkathm} 
\newtheorem{Thm}{Theorem}
\theoremstyle{bkathm} 
\newtheorem{Cor}{Corollary}
\theoremstyle{bkathm} 
\newtheorem{Lem}{Lemma}
\theoremstyle{definition}

\begin{document}
\setstretch{1.3}
\title{On the asymptotic distribution of the symmetrized Chatterjee's correlation coefficient}
\author{\normalsize Qingyang Zhang\\
\normalsize Department of Mathematical Sciences, University of Arkansas, AR 72701\\
\normalsize Email: qz008@uark.edu
}
\date{}
\maketitle

\begin{abstract}
Chatterjee (2021) introduced an asymmetric correlation measure that has attracted much attention over the past year. In this paper, we derive the asymptotic distribution of the symmetric version of Chatterjee's correlation, and suggest a finite sample test for independence.
\end{abstract}


\section{Introduction}
Suppose $X$ and $Y$ are two continuous random scalars, and $\{(X_{1}, Y_{1}), ..., (X_{n}, Y_{n})\}$ are $n$ i.i.d. samples from $(X, Y)$. In this paper, we are interested in the following classical independence test
\begin{align*}
H_{0}&: X\perp Y,\\
H_{a}&: X \not\perp Y.
\end{align*}

The problem of testing independence has been examined from a number of perspectives. For instance, Pfister et al. (2018) introduced a kernel-based test for joint independence between multiple variables \cite{hsic}. In their seminal work, Sz\'ekely et al. proposed a distance-based measure to test the association between random vectors of arbitrary dimensions \cite{dcor}. Bergsma \& Dassios (2014) defined a consistent test based on the sign covariance which is closely related to Kendall's $\tau$ \cite{BD}. Schweizer \& Wolff (1981) used copulas to obtain several nonparametric measures of dependence that satisfy Renyi's conditions \cite{SW}. There are many other proposals including a graph-theoretic measure \cite{graph} and the maximal information coefficient \cite{mic}.  

Recently, Chatterjee (2021) introduced a novel correlation coefficient based on simple rank statistics \cite{chatterjee}. His test has quickly attracted much attention as it is distribution-free, consistent against all fixed alternatives, and asymptotically normal under independence. We begin with a brief review of this test. Suppose $X_{i}$'s and $Y_{i}$'s have no ties, there is a unique way to rearrange the data (with respect to $X$) as $(X_{(1)}, Y_{X(1)}), ..., (X_{(n)}, Y_{X(n)})$, where $X_{(1)}<\cdots<X_{(n)}$ and $\{Y_{X(1)}, ..., Y_{X(n)}\}$ denote the concomitants. Let $R(Y_{X(i)})$ be the rank of $Y_{X(i)}$, i.e., $R(Y_{X(i)}) = \sum_{j=1}^{n}\mathbbm{1}\{Y_{(j)}\leq Y_{X(i)}\}$, Chatterjee defined the following correlation coefficient 
\begin{equation}
\xi_{n}(X, Y) = 1-\frac{3}{n^{2}-1}\sum_{i=1}^{n-1}|R(Y_{X(i+1)})-R(Y_{X(i)})|, 
\end{equation}
and showed that $\xi_{n}(X, Y)$ consistently estimates Dette-Siburg-Stoimenov's correlation \cite{dss}
\begin{equation*}
\xi(X, Y) = \frac{\int V(E(\mathbbm{1}\{Y\geq t|X\}))dF_{Y}(t)}{\int V(\mathbbm{1}\{Y\geq t\})dF_{Y}(t)}.
\end{equation*}

It is not hard to see that the population quantity $\xi(X, Y)$ is between 0 and 1, since $V(\mathbbm{1}\{Y\geq t\})\geq V(E(\mathbbm{1}\{Y\geq t|X\}))$ for any $t$. To be specific, when $X$ and $Y$ are independent,
$E(\mathbbm{1}\{Y\geq t|X\})$ is a constant for any $t$, therefore $\xi(X, Y)=0$. When $Y$ is a measurable function of $X$, we have $E(\mathbbm{1}\{Y\geq t|X\}) = \mathbbm{1}\{Y\geq t|X\}$ thus $\xi(X, Y)=1$. The sample estimate in (1) is asymptotically normal under independence. Precisely, 
$$\sqrt{n}\xi_{n}(X, Y)\xrightarrow[]{d} N(0, 2/5)$$ 
as $n\rightarrow\infty$. Moreover, as an empirical finding, Chatterjee's test is more sensitive to oscillatory patterns compared to other popular tests including distance correlation test and Bergsma-Dassios test. These attractive properties make Chatterjee's test an appealing choice for large-scale association studies. 

It is also noteworthy that $\xi_{n}(X, Y)$ is generally asymmetric, i.e., $\xi_{n}(X, Y)\neq \xi_{n}(Y, X)$. A symmetrized measure can be easily obtained by taking the maximum 
$$\xi^{sym}_{n}(X, Y) = \max\{\xi_{n}(X, Y) , \xi_{n}(Y, X) \}.$$ 
Unfortunately, the asymptotic distribution of $\xi^{sym}_{n}(X, Y)$ still remains unknown, although Chatterjee conjectured that under independence, $\xi^{sym}_{n}(X, Y)$ behaves like the maximum of two correlated normal random variables \cite{chatterjee}. To fill this gap, in this work we investigate the asymptotic joint distribution of $\sqrt{n}\xi_{n}(X, Y)$ and $\sqrt{n}\xi_{n}(Y, X)$. We show that under independence, $[\sqrt{n}\xi_{n}(X, Y), \sqrt{n}\xi_{n}(Y, X)]^{T}$ converges to a bivariate normal distribution with $\rho = 0$, therefore the maximum converges to a skew normal distribution with shape parameter $1$. As a byproduct, we derive the finite sample variance and covariance of $\sqrt{n}\xi_{n}(X, Y)$ and $\sqrt{n}\xi_{n}(Y, X)$, and construct a finite sample test to reduce the bias of asymptotic p-value. 

The remainder of the paper is structured as follows: Section 2 presents our main results including the asymptotic joint distribution of $\sqrt{n}\xi_{n}(X, Y)$ and $\sqrt{n}\xi_{n}(Y, X)$, as well as their finite sample variance and covariance. Section 3 compares the asymptotic test and finite sample test using simulated data. Section 4 discusses our method with some future perspectives.

\section{Main results}
We first present the finite sample covariance between $\sqrt{n}\xi_{n}(X, Y)$ and $\sqrt{n}\xi_{n}(Y, X)$ in the following lemma. A detailed proof of Lemma 1 is provided in Appendix A.1.  

\begin{Lem}
If $X$ and $Y$ are independent, we have
\begin{equation*}
\mbox{Cov}\left[\sqrt{n}\xi_{n}(X, Y), \sqrt{n}\xi_{n}(Y, X)\right ] = -n + \frac{9}{(n+1)^2(n-1)^3}\left [2(n-1)^2+(n-2)(A+B)^2+(n-2)(n-3)C^2\right ],
\end{equation*}
for any $n\geq 4$, where $A$, $B$, $C$ represent 
\begin{align*}
A & = \frac{1}{2(n-2)}\sum_{i=1}^{n-1} \left [ i(i-1)+(n-i+2)(n-i-1) \right ], \\
B & = \frac{1}{2(n-2)}\sum_{i=1}^{n-1} \left[(i+2)(i-1)+(n-i-1)(n-i)\right], \\
C & = \frac{1}{3(n-2)(n-3)} \sum_{i=1}^{n-1}\left[ n(n-1)(n+1) - 6(n-i)^2 - 6(i+1)(i-1) \right ].
\end{align*}
\end{Lem}

Figure 1 sketches $\mbox{Cov}\left[\sqrt{n}\xi_{n}(X, Y), \sqrt{n}\xi_{n}(Y, X)\right ]$ for $4\leq n\leq 100$. The covariance function is maximized at $n=6$ and decreases to $0$ afterwards. To verify Lemma 1, we also estimated the covariance between $\sqrt{n}\xi_{n}(X, Y)$ and $\sqrt{n}\xi_{n}(Y, X)$ using $100,000$ Monte Carlo samples. The analytical calculations and simulation results agree very well (see the comparison in Figure 1). The following corollary can be obtained from Lemma 1 (proof is given in Appendix A.2)

\begin{Cor}
If $X$ and $Y$ are independent, $\mbox{Cov}\left[\sqrt{n}\xi_{n}(X, Y), \sqrt{n}\xi_{n}(Y, X)\right ] = O(1/n)$. 
\end{Cor}

Furthermore, we give the finite sample variance in Lemma 2 (derivation is provided in Appendix A.3). The variance function is sketched in Figure 2 and verified by $100,000$ Monte Carlo samples. As stated in Lemma 2, $\mbox{V}\left[\sqrt{n}\xi_{n}(X, Y)\right ]$ monotonically increases from $0$ to $2/5$ for $n\geq 2$, therefore using the asymptotic variance of $2/5$ tends to overestimate the p-value for independence testing.

\begin{Lem}
If $X$ and $Y$ are independent, we have
\begin{equation*}
\mbox{V}\left[\sqrt{n}\xi_{n}(X, Y)\right ] = \frac{n(n-2)(4n-7)}{10(n+1)(n-1)^2},
\end{equation*}
for any $n\geq 2$. Furthermore, $\mbox{V}\left[\sqrt{n}\xi_{n}(X, Y)\right ]$ is monotonically increasing and $\mbox{V}\left[\sqrt{n}\xi_{n}(X, Y)\right ] \rightarrow 2/5$ as $n\rightarrow \infty$.
\end{Lem}

By Corollary 1 and Lemma 2, it is immediate that the correlation between $\sqrt{n}\xi_{n}(X, Y)$ and $\sqrt{n}\xi_{n}(Y, X)$ converges to 0 as $n\rightarrow \infty$. As of now, we have $\sqrt{n}\xi_{n}(X, Y)\xrightarrow[]{d} N(0, 2/5)$, $\sqrt{n}\xi_{n}(Y, X)\xrightarrow[]{d} N(0, 2/5)$ and $\mbox{Cor}\left[\sqrt{n}\xi_{n}(X, Y), \sqrt{n}\xi_{n}(Y, X)\right ] \rightarrow 0$, however, these do not imply asymptotic joint normality. The following lemma establishes the asymptotic joint normality using a coupling method by \cite{angus}, and Chatterjee's central limit theorem \cite{chatterjee.clt, auddy} (proof is given in Appendix A.4)  

\begin{Lem}
If $X$ and $Y$ are independent, $\left [ \sqrt{n}\xi_{n}(X, Y), \sqrt{n}\xi_{n}(Y, X)\right]$ converges in distribution to a bivariate normal distribution.
\end{Lem}

Combining Corollary 1, Lemma 2 and Lemma 3, our main theorem follows immediately

\begin{Thm}
If $X$ and $Y$ are independent, $\begin{bmatrix}
\sqrt{n}\xi_{n}(X, Y)\\
\sqrt{n}\xi_{n}(Y, X) 
\end{bmatrix}
\xrightarrow{d} N\left [
\begin{pmatrix}
0\\
0 
\end{pmatrix}, 
\begin{pmatrix}
2/5 & 0\\
0 & 2/5
\end{pmatrix} \right ]$ as $n\rightarrow\infty$.
\end{Thm}

It is well known that the maximum of two correlated standard normal random variables has a skew normal distribution with mean 0, variance 1 and shape $(1-\rho)/\sqrt{1-\rho^2}$, therefore we have the following asymptotic result for the symmetric measure  
\begin{equation*}
\sqrt{n}\xi^{sym}_{n}(X, Y) \rightarrow \mbox{SN}(\mu = 0, \sigma^2 = 2/5, \alpha = 1),
\end{equation*}
where $\mu$, $\sigma^2$ and $\alpha$ represent the location, variance and shape parameters in the skew normal distribution. 

Going back to Lemma 1 and Lemma 2, we note that a finite sample test for $\sqrt{n}\xi^{sym}_{n}(X, Y)$ may help reduce the bias of asymptotic p-value, and this can be done by replacing the asymptotic variance and shape parameters with the corresponding finite sample quantities, i.e.,
$$\mbox{SN}\left\{\mu = 0, \sigma^2 = \mbox{V}\left [\sqrt{n}\xi_{n}(X, Y)\right ], \alpha = \frac{1-\rho_{n}}{\sqrt{1-\rho_{n}^2}}\right\},$$
where 
$$\rho_{n} = \frac{\mbox{Cov}\left[\sqrt{n}\xi_{n}(X, Y), \sqrt{n}\xi_{n}(Y, X)\right ] }{\sqrt{\mbox{V}\left [\sqrt{n}\xi_{n}(X, Y)\right ]}\sqrt{ \mbox{V}\left [\sqrt{n}\xi_{n}(Y, X)\right ]}}.$$

\noindent For instance, when $n=10$, the finite sample method uses variance $0.296$ and shape parameter $0.637$, compared to $0.4$ and $1$ in the asymptotic method. In Section 3, we shall use simulations to evaluate the bias reduction under relatively small sample sizes.

\section{Simulation study}
In this section, two simulation studies were conducted to investigate the bias of p-values from the asymptotic and finite sample tests. The exact p-values were approximated using $5,000$ permutations and the bias was computed as the p-value by each test minus the exact p-value. Sample size was set to be $\{10, 20, 30, 40, 50\}$. 

In the first study, we examined the p-value bias under independence. In each simulation run, we generated $X$ from $\mbox{Uniform}[-1, 1]$ and $Y$ from $N(0,1)$ independently, and then computed the p-value bias of the two tests. Figure 3 summarizes the results over $1,000$ simulations runs. It can be seen that the asymptotic p-value is heavily biased for smaller sample sizes, but the use of finite sample test can significantly reduce the bias. For instance, when $n=10$, the median bias of the two tests are $0.11$ (asymptotic) and $0.03$ (finite sample). 

In the second study, we investigated the p-value bias under various dependence settings. Generating $X$ from $\mbox{Uniform}[-1, 1]$, the following four alternatives were considered
\begin{itemize}
\item[1.] Linear: $Y = X+\epsilon$, where $\epsilon\perp X$ and $\epsilon\sim N(0,1)$. 
\item[2.] Quadratic: $Y=X^2+\epsilon$.
\item[3.] Sinusoid: $Y = \cos(2\pi X)+2\epsilon$.
\item[4.] W-shaped: $Y = |X+0.5|\mathbbm{1}\{X<0\}+|X-0.5|\mathbbm{1}\{X>0\}+\epsilon$.
\end{itemize}

Figure 4 shows the bias comparison under different scenarios, where it is clear that the finite sample test again substantially reduces the bias. The results also indicate that the asymptotic test can be conservative and subsequently less powerful due to the positive bias. Therefore in practice, the finite sample test is recommended especially for small sample sizes. 

\section{Discussion and conclusions}
In a recent work, Chatterjee proposed an ingenious correlation measure that has several unusual appeals. For instance, it consistently estimates a population quantity that is 0 under independence and 1 under deterministic relation. Moreover, it is distribution-free and asymptotic normal under independence. Despite the success of this correlation measure, there are several  important problems remaining unexplored, one of which is about the asymptotic distribution of the symmetrized measure $\xi^{sym}_{n}(X, Y) = \max\{\xi_{n}(X, Y) , \xi_{n}(Y, X)\}$. In this paper, we showed that $\sqrt{n}\xi_{n}(X, Y)$ and $\sqrt{n}\xi_{n}(Y, X)$ are asymptotically bivariate normal with $\rho = 0$, thus $\sqrt{n}\xi^{sym}_{n}(X, Y)$ converges to a skew normal distribution with shape parameter 1. Motivated by the variance and covariance of $\sqrt{n}\xi_{n}(X, Y)$ and $\sqrt{n}\xi_{n}(Y, X)$, we also suggest a finite sample test for small sample sizes (e.g., $n<30$). Simulation studies show that the asymptotic test tends to be heavily biased, and the use of our finite sample test can reduce the bias.
 
Our results can be very useful for large-scale exploratory analyses in which the dependence of interest are non-directional. For example, in gene co-expression analysis, one may simply search for all the correlated gene pairs without specifying the direction, thus a symmetric measure of dependence is more appropriate. It is also noteworthy that, our finite sample test is computationally efficient because the variance and covariance terms only depend on the sample size $n$ (see Lemma 1 and Lemma 3).   

More recently, several studies have pointed out that Chatterjee's independence test is unfortunately sub-optimal \cite{ShiHan,LinHan}. As shown in Figure 5 of \cite{chatterjee}, Chatterjee's test has inferior performance for smoother alternatives, and its statistical power quickly deteriorates as the noise level increases. To boost the power of Chatterjee's test, Lin \& Han (2021) proposed the following extension of Chatterjee's original proposal by incorporating $M$ right nearest neighbors
\begin{equation*}
\xi_{n, M}(X, Y) = -2+\frac{6\sum_{i=1}^{n}\sum_{m=1}^{M}\min[R(Y_{X(i)}), R(Y_{X(i+m)})]}{(n+1)[nM+M(M+1)/4]},
\end{equation*}
where $R(Y_{X(i+m)})=R(Y_{X(i)})$ for $i+m>n$. Under the assumption that $M = o(n^{1/4})$, Lin \& Han showed $\sqrt{nM}\xi_{n, M}(X, Y) \xrightarrow[]{d} N(0, 2/5)$ as $n\rightarrow \infty$. Empirical data analyses suggest that, under a reasonable choice of $M$, $\xi_{n, M}(X, Y)$ achieves better statistical power than $\xi_{n}(X, Y)$ as the latter does not use any information from neighboring points. 

Similar to what we did in this work, it would be of interest to investigate the joint distribution of $\sqrt{nM}\xi_{n, M}(X, Y)$ and $\sqrt{nM}\xi_{n, M}(Y, X)$ and construct asymptotic and finite sample tests for the symmetric measure. However, the derivations for $M\geq 2$ may take significant effort and we leave this for future research.

 \section*{Competing Interests}
\noindent
The author has declared that no competing interests exist.

\newpage
\section*{Figures}
\begin{figure}[!htbp]
\begin{center}
\includegraphics[scale=0.35]{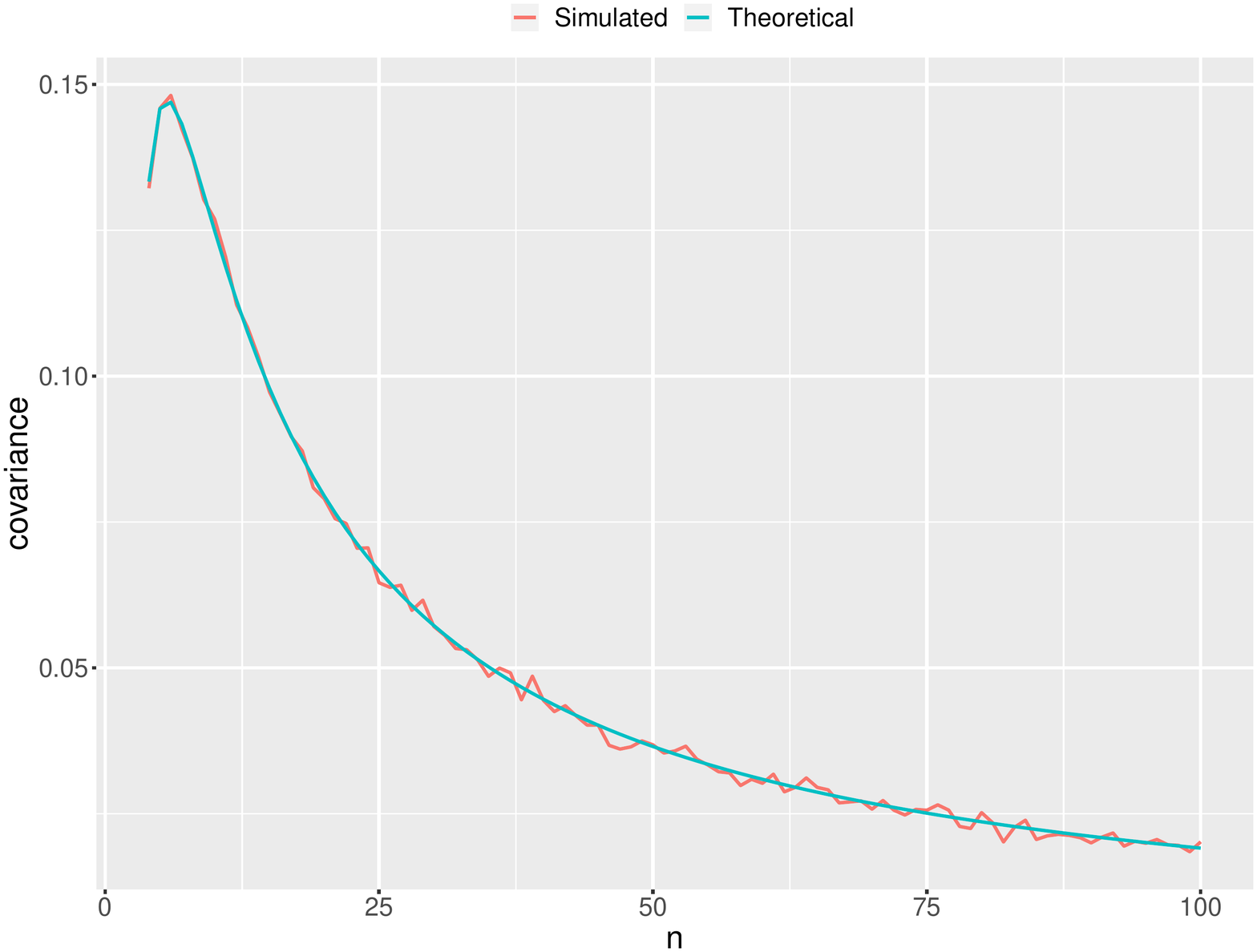}
\end{center}
\caption{The covariance between $\sqrt{n}\xi_{n}(X, Y)$ and $\sqrt{n}\xi_{n}(Y, X)$ based on theoretical calculation (Lemma 1) and 100,000 Monte Carlo samples.
}
\end{figure}

\newpage
\begin{figure}[!htbp]
\begin{center}
\includegraphics[scale=0.38]{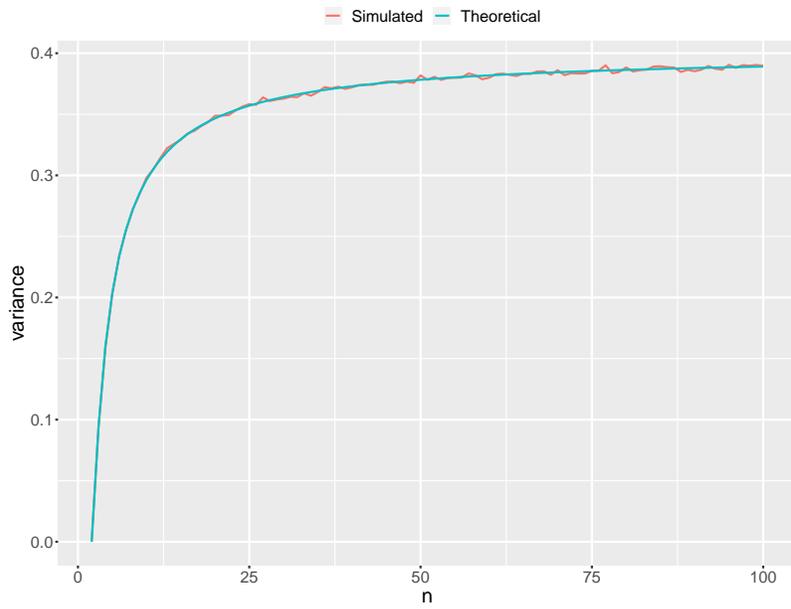}
\end{center}
\caption{The variance of $\sqrt{n}\xi_{n}(X, Y)$ based on theoretical calculation (Lemma 2) and 100,000 Monte Carlo samples.
}
\end{figure}

\newpage
\begin{figure}[!htbp]
\begin{center}
\includegraphics[scale=0.4]{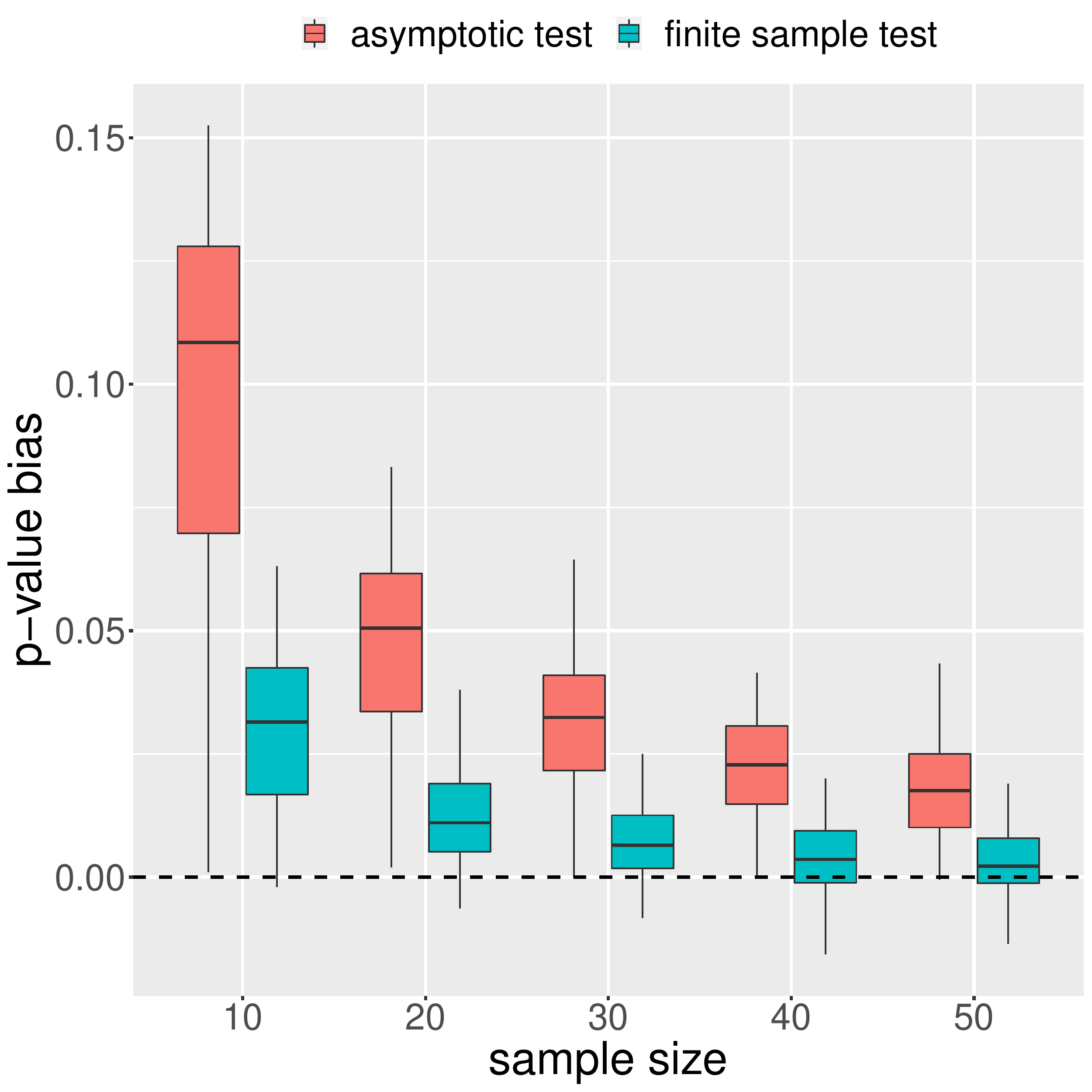}
\end{center}
\caption{Bias comparison of the asymptotic and finite sample p-values under independence. The bias is computed as the p-value by each test minus the exact p-value based on $5,000$ permutations.
}
\end{figure}

\newpage
\begin{figure}[!htbp]
\begin{center}
\includegraphics[scale=0.55]{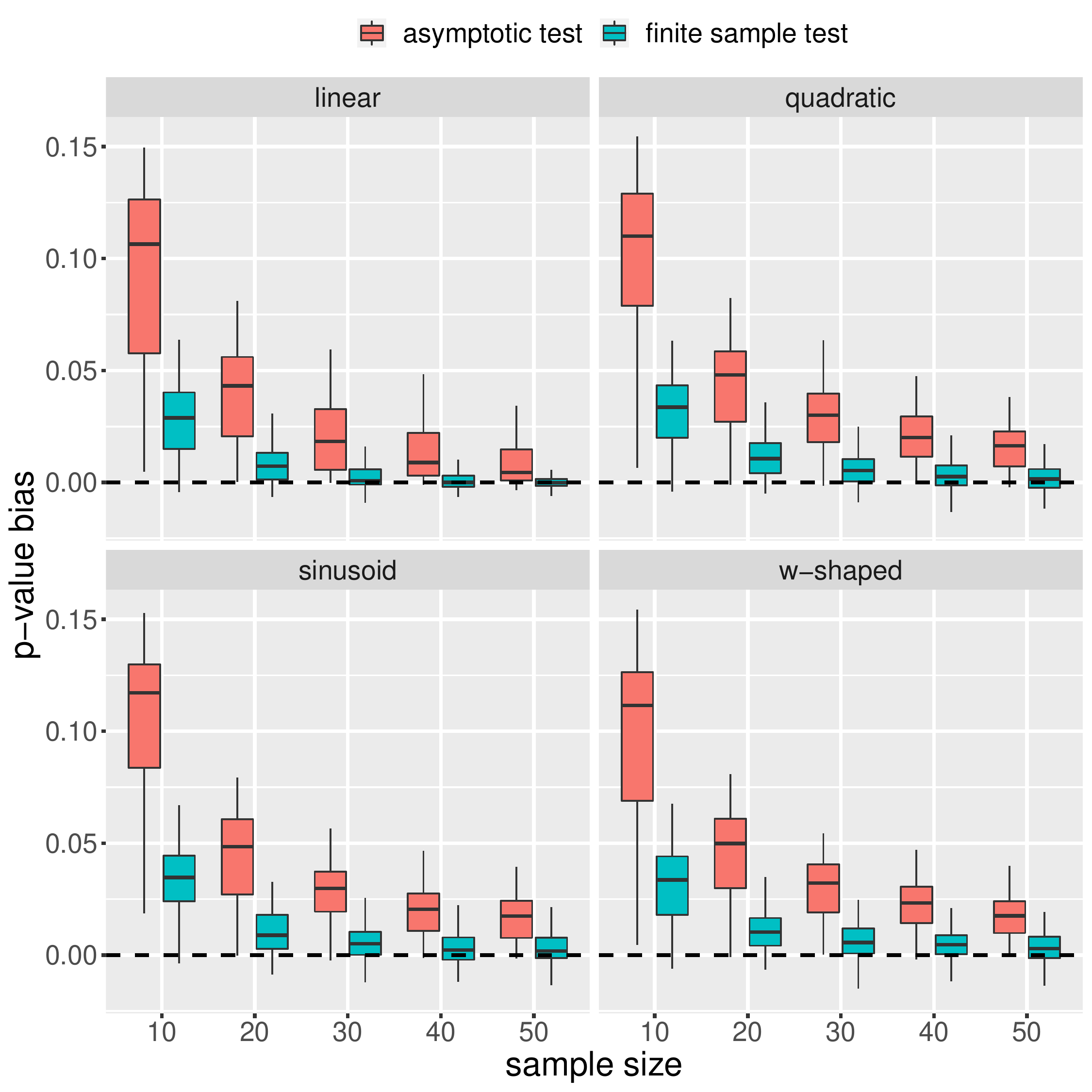}
\end{center}
\caption{Bias comparison of the asymptotic and finite sample p-values under four alternatives. The bias is computed as the p-value by each test minus the exact p-value based on $5,000$ permutations.
}
\end{figure}

\newpage
\section*{Appendix}
\subsection*{A.1. Proof of Lemma 1}
We first derive $E\left[|R(X_{Y(j+1)})-R(X_{Y(j)})||R(Y_{X(i+1)})-R(Y_{X(i)})|\right]$ using the law of total expectation. The space of $\left[ R(Y_{X(i+1)}), R(Y_{X(i)}) \right]$ can be partitioned into the following parts
\begin{align*}
Z_{1} & = \{R(Y_{X(i)}) = j, R(Y_{X(i)}) = j+1\} \cup \{R(Y_{X(i)}) = j+1, R(Y_{X(i+1)}) = j\},\\
Z_{2} & = \{R(Y_{X(i)}) = j, R(Y_{X(i+1)}) \neq (j, j+1)\}, \\
Z_{3} & = \{R(Y_{X(i+1)}) = j, R(Y_{X(i)}) \neq (j, j+1)\},\\
Z_{4} & = \{R(Y_{X(i)}) = j+1, R(Y_{X(i+1)}) \neq (j, j+1)\}, \\
Z_{5} & = \{R(Y_{X(i+1)}) = j+1, R(Y_{X(i)}) \neq (j, j+1)\},\\
Z_{6} & = \{R(Y_{X(i)}) \neq (j, j+1), R(Y_{X(i+1)}) \neq (j, j+1)\},
\end{align*}
It is straightforward that $|R(X_{Y(j+1)})-R(X_{Y(j)})|$ and $|R(Y_{X(i+1)})-R(Y_{X(i)})|$ are independent given each $Z_{m}, m=1, ..., 6$. The probabilities of $Z_{m}$ are
\begin{align*}
P(Z_{1}) & = \frac{2}{n(n-1)},\\
P(Z_{2}) & = P(Z_{3}) = P(Z_{4}) = P(Z_{5}) =\frac{n-2}{n(n-1)},\\
P(Z_{6}) & =  \frac{(n-2)(n-3)}{n(n-1)},
\end{align*}
and the following expectations can be derived using elementary probability
\begin{align*}
E\left[[|R(X_{Y(j+1)})-R(X_{Y(j)})||Z_{1}\right] &= E\left|R(Y_{X(i+1)})-R(Y_{X(i)})||Z1\right] =1,  \\
E\left[|R(X_{Y(j+1)})-R(X_{Y(j)})||Z_{2}\right] & = \frac{(i-1)i+(n-i+2)(n-i-1)}{2(n-2)}, \\
E\left[|R(Y_{X(i+1)})-R(Y_{X(i)})||Z_{2}\right] & = \frac{(j-1)j+(n-j+2)(n-j-1)}{2(n-2)}, \\
E\left[|R(X_{Y(j+1)})-R(X_{Y(j)})||Z_{3}\right] & = \frac{(i+2)(i-1)+(n-i-1)(n-i)}{2(n-2)}, \\
E\left[|R(Y_{X(i+1)})-R(Y_{X(i)})||Z_{3}\right] & = \frac{(j-1)j+(n-j+2)(n-j-1)}{2(n-2)}, 
\end{align*}
\begin{align*}
E\left[|R(X_{Y(j+1)})-R(X_{Y(j)})||Z_{4}\right] & = \frac{(i-1)i+(n-i+2)(n-i-1)}{2(n-2)}, \\
E\left[|R(Y_{X(i+1)})-R(Y_{X(i)})||Z_{4}\right] & = \frac{(j+2)(j-1)+(n-j-1)(n-j)}{2(n-2)}, \\
E\left[|R(X_{Y(j+1)})-R(X_{Y(j)})||Z_{5}\right] & = \frac{(i+2)(i-1)+(n-i-1)(n-i)}{2(n-2)}, \\
E\left[|R(Y_{X(i+1)})-R(Y_{X(i)})||Z_{5}\right] & = \frac{(j+2)(j-1)+(n-j-1)(n-j)}{2(n-2)}, \\
E\left[|R(X_{Y(j+1)})-R(X_{Y(j)})||Z_{6}\right] & = \frac{n(n-1)(n+1) - 6(n-i)^2 - 6(i+1)(i-1)}{3(n-2)(n-3)}, \\
E\left[|R(Y_{X(i+1)})-R(Y_{X(i)})||Z_{6}\right] & = \frac{n(n-1)(n+1) - 6(n-j)^2 - 6(j+1)(j-1)}{3(n-2)(n-3)}
\end{align*}
For the ease of notation, let 
\begin{align*}
A_{i} & = \frac{(i-1)i+(n-i+2)(n-i-1)}{2(n-2)},\\
B_{i} & = \frac{(i+2)(i-1)+(n-i-1)(n-i)}{2(n-2)},\\
C_{i} & = \frac{n(n-1)(n+1) - 6(n-i)^2 - 6(i+1)(i-1)}{3(n-2)(n-3)},
\end{align*}
then by the law of total expectation and conditional independence between $|R(X_{Y(j+1)})-R(X_{Y(j)})|$ and $|R(Y_{X(i+1)})-R(Y_{X(i)})|$, we have
\begin{align*}
& E\left[|R(X_{Y(j+1)})-R(X_{Y(j)})||R(Y_{X(i+1)})-R(Y_{X(i)})|\right] \\
= &\sum_{m=1}^{6}E\left[|R(X_{Y(j+1)})-R(X_{Y(j)})||Z_{m}\right]E\left[|R(Y_{X(i+1)})-R(Y_{X(i)})||Z_{m}\right]P(Z_{m}) \\
= &\frac{2}{n(n-1)}+\frac{n-2}{n(n-1)}(A_{i}A_{j}+B_{i}B_{j}+A_{i}B_{j}+A_{j}B_{i})+\frac{(n-2)(n-3)}{n(n-1)}C_{i}C_{j}.
\end{align*}
Using the following fact
\begin{equation*}
E\left[|R(X_{Y(j+1)})-R(X_{Y(j)})|\right] E\left[|R(Y_{X(i+1)})-R(Y_{X(i)})|\right] = \frac{(n+1)^2}{9},
\end{equation*}
the covariance can be obtained immediately
\begin{align*}
& \mbox{Cov}\left[|R(X_{Y(j+1)})-R(X_{Y(j)})|, |R(Y_{X(i+1)})-R(Y_{X(i)})|\right] \\
= & \frac{2}{n(n-1)}+\frac{n-2}{n(n-1)}(A_{i}+B_{i})(A_{j}+B_{j})+\frac{(n-2)(n-3)}{n(n-1)}C_{i}C_{j}-\frac{(n+1)^2}{9}.
\end{align*}
Taking the summation over $i$ and $j$, Lemma 1 is proved.

\subsection*{A.2. Proof of Corollary 1}
Using $\sum_{i=1}^{n-1}i^2 = \sum_{i=1}^{n-1}(n-i)^2 = n(n-1)(2n-1)/6$, we have
\begin{align*}
A & = \frac{1}{2(n-2)}\left[ \frac{n(n-1)(2n-1)}{3} + O(n^2) \right] \\
& = \frac{n^2}{3}+ O(n)
\end{align*}
Similarly, $B = n^2/3 + O(n)$, therefore $(n-2)(A+B)^2 = 4n^5/9 + O(n^4)$. Furthermore,
\begin{align*}
(n-2)(n-3)C^{2} & = \frac{1}{9(n-2)(n-3)}\left[ n(n-1)^{2}(n+1) - 2n(n-1)(2n-1) + O(n) \right]^2 \\
 & = \frac{n^{2}(n-1)^2}{9(n-2)(n-3)}\left[ n^2-4n+1+O(1/n)\right]^2 \\
 & = \frac{n^{4}(n-1)^{2}(n-4)^{2}}{9(n-2)(n-3)}+O(n^4)\\
 & = \frac{n^6}{9} - \frac{5n^5}{9} + O(n^4).
\end{align*}
Summarizing the results above, we have 
\begin{align*}
\mbox{Cov}\left[\sqrt{n}\xi_{n}(X, Y), \sqrt{n}\xi_{n}(Y, X)\right ] & = -n + \frac{n^5}{(n+1)^{2}(n-1)^{2}} + O(1/n) \\
& = \frac{2n^3-n}{(n+1)^{2}(n-1)^{2}} + O(1/n) \\
& = O(1/n)
\end{align*}

\subsection*{A.3. Proof of Lemma 2}
If $X$ and $Y$ are independent, $\{R(Y_{X(1)}), ..., R(Y_{X(n)})\}$ is a random permutation of $\{1, ..., n\}$. For the ease of notation, we let $R_{i} = R(Y_{X(i)})$. The following facts from \cite{LinHan} (Lemma 6.1, page 13) are needed for our derivation of $\mbox{V}\left[\sqrt{n}\xi_{n}(X, Y)\right ]$
\begin{align*}
\mbox{V}\left[ R_{1} \right] & = \frac{(n-1)(n+1)}{12} \\
\mbox{Cov}\left[ R_{1}, R_{2}\right] & = -\frac{n+1}{12} \\
\mbox{Cov}\left[ R_{1}, \min\{R_{2}, R_{3}\}\right] & = -\frac{n+1}{12} \\
\mbox{Cov}\left[ R_{1}, \min\{R_{1}, R_{2}\}\right] & = \frac{(n+1)(n-2)}{24} \\
\mbox{Cov}\left[ \min\{R_{1}, R_{2}\}, \min\{R_{3}, R_{4}\}\right] & = -\frac{4(n+1)}{45} \\
\mbox{Cov}\left[ \min\{R_{1}, R_{2}\}, \min\{R_{1}, R_{3}\}\right] & = \frac{(n+1)(4n-17)}{180} \\
\mbox{V}\left[ \min\{R_{1}, R_{2}\} \right] & = \frac{(n-2)(n+1)}{18} .
\end{align*}
We first decompose the total variance into three parts
\begin{align*}
\mbox{V}\left[ \sum_{i=1}^{n-1}|R_{i+1}-R_{i}| \right] = & (n-1)\mbox{V}\left[|R_{2}-R_{1}|\right] + 2(n-2)\mbox{Cov}\left[ |R_{2}-R_{1}|, |R_{3}-R_{2}|\right]\\
&+ (n-2)(n-3)\mbox{Cov}\left[ |R_{4}-R_{3}|, |R_{2}-R_{1}|\right]
\end{align*}
where $\mbox{V}\left[|R_{2}-R_{1}|\right]$ in the first part can be derived as follows
\begin{align*}
\mbox{V}\left[|R_{2}-R_{1}| \right]& = \mbox{V}\left[R_{1}+R_{2} - 2\min\{R_{1}, R_{2}\} \right]\\
& =  2\mbox{V}\left[R_{1}\right] + 2\mbox{Cov}\left[R_{1}, R_{2}\right] + 4\mbox{V}\left[ \min\{R_{1}, R_{2}\} \right] -8\mbox{Cov}\left[ R_{1}, \min\{R_{1}, R_{2}\}\right]\\
& = \frac{(n+1)(n-1)}{6} - \frac{n+1}{6} + \frac{2(n-2)(n+1)}{9} - \frac{(n-2)(n+1)}{3} \\
& = \frac{(n+1)(n-2)}{18},
\end{align*}
and $\mbox{Cov}\left[ |R_{2}-R_{1}|, |R_{3}-R_{2}|\right]$ in the second part is
\begin{align*}
& \mbox{Cov}\left[ |R_{2}-R_{1}|, |R_{3}-R_{2}|\right] = \mbox{Cov}\left[ R_{1}+R_{2}-2\min\{R_{1}, R_{2}\}, R_{2}+R_{3}-2\min\{R_{2}, R_{3}\} \right]\\
= &  3\mbox{Cov}\left[ R_{1}, R_{2}\right] +  \mbox{V}\left[ R_{1} \right] - 4\mbox{Cov}\left[ R_{1}, \min\{R_{2}, R_{3}\}\right] - 4\mbox{Cov}\left[ R_{1}, \min\{R_{1}, R_{2}\}\right] + 4\mbox{Cov}\left[ \min\{R_{1}, R_{2}\}, \min\{R_{1}, R_{3}\}\right] \\
= & -\frac{n+1}{4}+\frac{(n+1)(n-1)}{12}+\frac{n+1}{3}-\frac{(n-2)(n+1)}{6}+\frac{(n+1)(4n-17)}{45}\\
= & \frac{(n+1)(n-8)}{180},
\end{align*}
and $\mbox{Cov}\left[ |R_{4}-R_{3}|, |R_{2}-R_{1}|\right]$ in the third part is 
\begin{align*}
& \mbox{Cov}\left[ |R_{4}-R_{3}|, |R_{2}-R_{1}|\right] = \mbox{Cov}\left[ R_{1}+R_{2}-2\min\{R_{1}, R_{2}\}, R_{3}+R_{4}-2\min\{R_{3}, R_{4}\} \right]\\
= &  4\mbox{Cov}\left[ R_{1}, R_{2}\right] - 8\mbox{Cov}\left[ R_{1}, \min\{R_{2}, R_{3}\}\right] + 4 \mbox{Cov}\left[ \min\{R_{1}, R_{2}\}, \min\{R_{3}, R_{4}\}\right] \\
= & -\frac{n+1}{3} + \frac{2(n+1)}{3}-\frac{16(n+1)}{45} \\
= & - \frac{n+1}{45}.
\end{align*}
Summarizing the results above, we have 
\begin{equation*}
\mbox{V}\left[\sqrt{n}\xi_{n}(X, Y)\right ] = \frac{n(n-2)(4n-7)}{10(n+1)(n-1)^2},
\end{equation*}
for any $n\geq 2$, and $\mbox{V}\left[\sqrt{n}\xi_{n}(X, Y)\right ] \rightarrow 2/5$ as $n\rightarrow \infty$. It is also noteworthy that 
\begin{equation*}
\mbox{V}\left[\sqrt{n}\xi_{n}(X, Y)\right ] = \frac{2}{5}(1-\frac{1}{n+1})(1-\frac{1}{n-1})(1-\frac{3/4}{n-1}),
\end{equation*}
thus $\mbox{V}\left[\sqrt{n}\xi_{n}(X, Y)\right ]$ is monotonically increasing.

\subsection*{A.4. Proof of Lemma 3}
For the ease of notation, let 
\begin{align*}
Z_{n}^{X} & = \frac{\sum_{i=1}^{n-1}|R(X_{Y(i+1)})-R(X_{Y(i)})| - n(n-1)/3}{\sqrt{n}(n-1)}, \\
Z_{n}^{Y} & =  \frac{\sum_{i=1}^{n-1}|R(Y_{X(i+1)})-R(Y_{X(i)})| - n(n-1)/3}{\sqrt{n}(n-1)}.
\end{align*}
Let $F_{X}(x)$ and $F_{Y}(y)$ be the $c.d.f.$'s of $X$ and $Y$, $\hat{F}_{X}(x)$ and $\hat{F}_{Y}(y)$ be the empirical $c.d.f.$'s, i.e.,  
\begin{align*}
\hat{F}_{X}(x) & =\frac{1}{n}\sum_{i=1}^{n}\mathbbm{1}\{X_{i}\leq x\}, \\
\hat{F}_{Y}(y) & =\frac{1}{n}\sum_{i=1}^{n}\mathbbm{1}\{Y_{i}\leq y\},
\end{align*}
therefore $\hat{F}_{X}(X_{Y(i)})=R(X_{Y(i)})/n$ and $\hat{F}_{Y}(Y_{X(i)}) = R(Y_{X(i)})/n$. We define $U_{i} := F_{X}(X_{Y(i)})$ and $V_{i} := F_{Y}(Y_{X(i)})$. Under independence, $\{U_{1}, ..., U_{n}\}$ are $i.i.d.$ samples from $\mbox{Uniform}(0, 1)$, and same for $\{V_{1}, ..., V_{n}\}$. But it is noteworthy that $\{U_{1}, ..., U_{n}\}$ and $\{V_{1}, ..., V_{n}\}$ are not independent. 

Using Equations (5)-(8) in Angus (1995), we have 
\begin{align*}
Z_{n}^{X} = & \frac{1}{\sqrt{n}}\sum_{i=1}^{n-1}\left[ |U_{i+1}-U_{i}| + 2U_{i}(1-U_{i}) - \frac{2}{3}\right] + R_{n}^{X},\\
Z_{n}^{Y} = & \frac{1}{\sqrt{n}}\sum_{i=1}^{n-1}\left[ |V_{i+1}-V_{i}| + 2V_{i}(1-V_{i}) - \frac{2}{3}\right] + R_{n}^{Y},
\end{align*}
where $R_{n}^{X}\xrightarrow{P}0$ and $R_{n}^{Y}\xrightarrow{P}0$. In addition, it can be seen that
\begin{equation*}
\sqrt{n}\xi_{n}(X, Y) = Z_{n}^{X} - \frac{\sum_{i=1}^{n-1}|R(Y_{X(i+1)})-R(Y_{X(i)})|}{\sqrt{n}(n-1)(n+1)},
\end{equation*}
where the second term is $O(1/\sqrt{n})$. For any two constants $a$ and $b$, we have
\begin{align*}
& a\sqrt{n}\xi_{n}(X, Y)+b\sqrt{n}\xi_{n}(Y, X) \\
= ~& aZ_{n}^{X} + bZ_{n}^{Y}+O(1/\sqrt{n}) \\
= ~ & \frac{1}{\sqrt{n}}\sum_{i=1}^{n-1}\left[ a|U_{i+1}-U_{i}| + b|V_{i+1}-V_{i}| + 2aU_{i}(1-U_{i}) + 2bV_{i}(1-V_{i}) - \frac{2(a+b)}{3}\right] \\
&~ + aR_{n}^{X} + bR_{n}^{Y} + O(1/\sqrt{n}).
\end{align*}

Recall that $U_{i} = F_{X}(X_{Y(i)})$ and $V_{i} = F_{Y}(Y_{X(i)})$, we will show that 
\begin{align}
\frac{1}{\sqrt{n}}\sum_{i=1}^{n-1}[ &a|F_{X}(X_{Y(i+1)})-F_{X}(X_{Y(i)})| + 2aF_{X}(X_{Y(i)})(1-F_{X}(X_{Y(i)})) \\
 &+ b|F_{Y}(Y_{X(i+1)})-F_{Y}(Y_{X(i)})| + 2bF_{Y}(Y_{X(i)})(1-F_{Y}(Y_{X(i)})) \nonumber \\
 &- 2(a+b)/3] \nonumber 
\end{align}
converges to a normal distribution. It is equivalent to show
\begin{align}
W_{n} = \frac{1}{\sqrt{n}}\sum_{i=1}^{n}[ &a|F_{X}(X_{i})-F_{X}(X_{N_{Y}(i)})| + 2aF_{X}(X_{i})(1-F_{X}(X_{i})) \\
 &+ b|F_{Y}(Y_{i})-F_{Y}(Y_{N_{X}(i)})| + 2bF_{Y}(Y_{i})(1-F_{Y}(Y_{i}))\nonumber  \\
 &- 2(a+b)/3],\nonumber 
\end{align}
converges to a normal distribution, where $\{X_{N_{X}(i)}, Y_{N_{X}(i)}\}$ and $\{X_{N_{Y}(i)}, Y_{N_{Y}(i)}\}$ represent the right nearest neighbor of $\{X_{i}, Y_{i}\}$ in terms of $X$ and $Y$, respectively. If $N_{X}(i)$ does not exist, one can define $Y_{N_{X}(i)} = Y_{(1)}$. Note that (3) is a summation over $i = 1, ..., n$, which has one more term than (2), but the two differ by only $O(1/\sqrt{n})$. 

It is easy to see that (3) is based on $n$ dependent variables, therefore the central limit theorem for $i.i.d.$ case does not apply. Here we use Chatterjee's central limit theorem based on interaction graphs \cite{chatterjee.clt, auddy}. We define a graphical rule $\mathcal{G}$ based on $W_{n}$ which we will show to be an interaction rule (defined in Section 2.3 of \cite{chatterjee.clt}, page 5). We borrow the notations from \cite{auddy}. Let
\begin{equation*}
\mathcal{M} :=\{(X_{1}, Y_{1}), ..., (X_{n}, Y_{n})\} ~\mbox{and}~ \mathcal{M}' := \{(X'_{1}, Y'_{1}), ..., (X'_{n}, Y'_{n})\}
\end{equation*}
be two $i.i.d.$ samples of size $n$, and 
\begin{align*}
\mathcal{M}^{i} & :=\{(X_{1}, Y_{1}), ..., (X'_{i}, Y'_{i}), ..., (X_{n}, Y_{n})\}  \\
\mathcal{M}^{j} & :=\{(X_{1}, Y_{1}), ..., (X'_{j}, Y'_{j}), ..., (X_{n}, Y_{n})\}  \\
\mathcal{M}^{ij} & :=\{(X_{1}, Y_{1}), ..., (X'_{i}, Y'_{i}), ..., (X'_{j}, Y'_{j}), ..., (X_{n}, Y_{n})\} 
\end{align*}

Let $[n] := \{1, 2, ..., n\}$, we define a graphical rule $\mathcal{G}(\mathcal{M})$ on $[n]$ motivated by \cite{auddy} (Equation 4.17, page 21)
$$
D^{X}_{\mathcal{M}}(i, j)=
\begin{cases}
\infty, ~~\mbox{if} ~X_{i}>X_{j}\\
\#\{l: X_{i}<X_{l}<X_{j}\}, ~~\mbox{if}~X_{i}<X_{j},
\end{cases}
$$
$$
D^{Y}_{\mathcal{M}}(i, j)=
\begin{cases}
\infty, ~~\mbox{if} ~Y_{i}>Y_{j}\\
\#\{l: Y_{i}<Y_{l}<Y_{j}\}, ~~\mbox{if}~Y_{i}<Y_{j}.
\end{cases}
$$
For any two indices $\{i, j\}$, there is an edge between $i$ and $j$ if there exists an $l \in [n]$, such that
\begin{align*}
D^{X}_{\mathcal{M}}(l, i)\leq 2~&\mbox{and}~D^{X}_{\mathcal{M}}(l, j)\leq 2 \\
& \mbox{or}\\
D^{Y}_{\mathcal{M}}(l, i)\leq 2~&\mbox{and}~D^{Y}_{\mathcal{M}}(l, j)\leq 2.
\end{align*}
It is straightforward that this rule is invariant under relabeling of indices, therefore it is a symmetric rule (see the definition in \cite{chatterjee.clt}, page 5). The statistic $W_{n}(\mathcal{M})$ can be separated into the $X$ part and the $Y$ part, i.e.,
$$W_{n}(\mathcal{M}) = W_{n, X}(\mathcal{M}) + W_{n, Y}(\mathcal{M}),$$
$$W_{n, X}(\mathcal{M}) = \frac{1}{\sqrt{n}}\sum_{i=1}^{n}[ a|F_{X}(X_{i})-F_{X}(X_{N_{Y}(i)})| + 2aF_{X}(X_{i})(1-F_{X}(X_{i})) - (a+b)/3],$$
$$W_{n, Y}(\mathcal{M}) = \frac{1}{\sqrt{n}}\sum_{i=1}^{n}[ b|F_{Y}(Y_{i})-F_{Y}(Y_{N_{X}(i)})| + 2bF_{Y}(Y_{i})(1-F_{Y}(Y_{(i)})) - (a+b)/3].$$

We will show that $\mathcal{G}(\mathcal{M})$ is an interaction rule. For any pair of indices $\{i, j\}$, if there is no edge between them, there does not exist an $l\in[n]$, such that $D^{X}_{\mathcal{M}}(l, i)\leq 2$ and $D^{X}_{\mathcal{M}}(l, j)\leq 2$. Using the proof of Lemma 4 in \cite{auddy} (page 43, section A.4), we have 
$$W_{n, X}(\mathcal{M})-W_{n, X}(\mathcal{M}^{i})-W_{n, X}(\mathcal{M}^{j})+W_{n, X}(\mathcal{M}^{ij}) = 0.$$
Similarly for the $Y$ part, we have
$$W_{n, Y}(\mathcal{M})-W_{n, Y}(\mathcal{M}^{i})-W_{n, Y}(\mathcal{M}^{j})+W_{n, Y}(\mathcal{M}^{ij}) = 0.$$
Therefore
$$W_{n}(\mathcal{M})-W_{n}(\mathcal{M}^{i})-W_{n}(\mathcal{M}^{j})+W_{n}(\mathcal{M}^{ij}) = 0,$$
i.e., $\mathcal{G}$ is a symmetric graphical interaction rule. Define
$$\Delta_{j} : = W_{n}(\mathcal{M})-W_{n}(\mathcal{M}^{j}),$$
and 
$$M := \max_{j}|\Delta_{j}|.$$
Because all the $c.d.f.$'s are between 0 and 1, there exists a constant $C_{1}>0$, such that 
$$M<\frac{C_{1}}{\sqrt{n}},$$
$$|\Delta_{j}|<\frac{C_{1}}{\sqrt{n}}.$$
We now construct an extended graph $\mathcal{G}'$ on $[n+4]$. There exists an edge between $i$ and $j$ if there exists an $l \in [n]$, such that
\begin{align*}
D^{X}_{\mathcal{M}}(l, i)\leq 6~&\mbox{and}~D^{X}_{\mathcal{M}}(l, j)\leq 6 \\
& \mbox{or}\\
D^{Y}_{\mathcal{M}}(l, i)\leq 6~&\mbox{and}~D^{Y}_{\mathcal{M}}(l, j)\leq 6.
\end{align*}
Clearly the extended rule $\mathcal{G}'$ is also symmetric, and all the edges of $\mathcal{G}$ are also in $\mathcal{G}'$. The degree of any vertex in $\mathcal{G}'$ is bounded by a constant $C_{2}>0$, therefore
$$\delta := 1+ \mbox{degree~of~the~vertex~1~in~} \mathcal{G}' \leq C_{2} + 1.$$
Using Theorem 2.5 of \cite{chatterjee.clt}, there exists a universal constant $C>0$, such that
$$\mathcal{D}(W_{n}) \leq \frac{C}{\sqrt{n}\sigma^2} + \frac{C}{2\sqrt{n}\sigma^3},$$
where $\mathcal{D}(W_{n})$ is the Wasserstein distance between $(W_{n}-E(W_{n}))/\sigma$ and $N(0 ,1)$.

We now derive the variance term $\sigma^{2}$. Note that 
\begin{align*}
\mbox{V}(W_{n})  = & \mbox{V}\left\{\frac{1}{\sqrt{n}}\sum_{i=1}^{n-1}[ a|U_{i+1}-U_{i}| + b|V_{i+1}-V_{i}| + 2aU_{i}(1-U_{i}) + 2bV_{i}(1-V_{i})])\right\}\\
 = & \mbox{V}\left\{\frac{1}{\sqrt{n}}\sum_{i=1}^{n-1}[ a|U_{i+1}-U_{i}| + 2aU_{i}(1-U_{i})]\right\} + \mbox{V}\left\{\frac{1}{\sqrt{n}}\sum_{i=1}^{n-1}[ b|V_{i+1}-V_{i}| + 2bV_{i}(1-V_{i})]\right\} \\
& + \mbox{Cov}\left\{\frac{1}{\sqrt{n}}\sum_{i=1}^{n-1}[ a|U_{i+1}-U_{i}| + 2aU_{i}(1-U_{i})], \frac{1}{\sqrt{n}}\sum_{i=1}^{n-1}[ b|V_{i+1}-V_{i}| + 2bV_{i}(1-V_{i})]\right\}.
\end{align*}
From Equation (14) in \cite{angus}, we have
\begin{align*}
\mbox{V}\left\{\frac{1}{\sqrt{n}}\sum_{i=1}^{n-1}[ a|U_{i+1}-U_{i}| + 2aU_{i}(1-U_{i})]\right\} & = 2a^2/45 + O(1/n) \\
\mbox{V}\left\{\frac{1}{\sqrt{n}}\sum_{i=1}^{n-1}[ b|V_{i+1}-V_{i}| + 2bV_{i}(1-V_{i})]\right\} & = 2b^2/45 + O(1/n) 
\end{align*}
For the covariance term, using the same technique in the proof of Lemma 1, we have
\begin{align*}
 \mbox{Cov}\left\{\frac{1}{\sqrt{n}}\sum_{i=1}^{n-1}[ a|U_{i+1}-U_{i}| ], \frac{1}{\sqrt{n}}\sum_{i=1}^{n-1}[ b|V_{i+1}-V_{i}| ]\right\} & =  O(1/n)\\
 \mbox{Cov}\left\{\frac{1}{\sqrt{n}}\sum_{i=1}^{n-1}[ 2aU_{i}(1-U_{i})], \frac{1}{\sqrt{n}}\sum_{i=1}^{n-1}[ 2bV_{i}(1-V_{i})]\right\} & = 0 \\
 \mbox{Cov}\left\{\frac{1}{\sqrt{n}}\sum_{i=1}^{n-1}[ a|U_{i+1}-U_{i}|], \frac{1}{\sqrt{n}}\sum_{i=1}^{n-1}[2bV_{i}(1-V_{i})]\right\} & = 0 \\
 \mbox{Cov}\left\{\frac{1}{\sqrt{n}}\sum_{i=1}^{n-1}[ 2aU_{i}(1-U_{i})], \frac{1}{\sqrt{n}}\sum_{i=1}^{n-1}[ b|V_{i+1}-V_{i}| ]\right\} & = 0.
\end{align*}
Summarizing the results above, we have
$$ \mbox{Cov}\left\{\frac{1}{\sqrt{n}}\sum_{i=1}^{n-1}[ a|U_{i+1}-U_{i}| + 2aU_{i}(1-U_{i})], \frac{1}{\sqrt{n}}\sum_{i=1}^{n-1}[ b|V_{i+1}-V_{i}| + 2bV_{i}(1-V_{i})]\right\} = O(1/n),$$
therefore
$$\sigma^2 := \mbox{Var}(W_{n})  = 2(a^2+b^2)/45 + O(1/n).$$

By Slutsky's theorem, $a\sqrt{n}\xi_{n}(X, Y)+b\sqrt{n}\xi_{n}(Y, X)$ converges in distribution to a normal distribution. Finally, by Cra\'mer-Wold device, $\left [\sqrt{n}\xi_{n}(X, Y), \sqrt{n}\xi_{n}(Y, X)\right ]^T$ is asymptotically bivariate normal.
 
\end{document}